\def\nii{[N~{\sc ii}]}
\def\teoiii{T$_{\rm e}$[O III]}
\def\tenii{T$_{\rm e}$[N II]}

\def\te{T$_{\rm e}}
\def\oiii{[O~{\sc iii}]}

\def\oii{[O~{\sc ii}]}
\def\ha{H$\alpha$}
\def\hb{H$\beta$}
\def\hg{H$\gamma$}

\def\hii{H~{\sc ii}}
\def\heii{He~{\sc ii}}

\def\hei{He~{\sc i}}
\def\sii{[S~{\sc ii}]}
\def\siii{[S~{\sc iii}]}
\def\cbeta{c({H$\beta$})}

\def\nii{[N~{\sc ii}]}

\def\oiii{[O~{\sc iii}]}

\def\oii{[O~{\sc ii}]}
\def\ha{H$\alpha$}
\def\hb{H$\beta$}
\def\hg{H$\gamma$}

\def\hii{H~{\sc ii}}
\def\hi{H~{\sc i}} 
\def\heii{He~{\sc ii}}

\def\hei{He~{\sc i}}
\def\sii{[S~{\sc ii}]}
\def\siii{[S~{\sc iii}]}
\def\te{$T_{e}$}

\documentclass[structabstract]{aa}  
\usepackage{graphicx}
\usepackage{rotating}
\usepackage{txfonts}
\usepackage{longtable,lscape}
\begin{document}

\title{The population of planetary nebulae and \hii~regions in M81.}
\subtitle{A study of radial metallicity gradients and chemical evolution.}

\author{Letizia Stanghellini\inst{1}
\and Laura Magrini\inst{2}
\and Eva Villaver\inst{3}
\and Daniele Galli\inst{2}
}
 
\institute{
National Optical Astronomy Observatories, Tucson, AZ 85719;
\email{lstanghellini@noao.edu}
\and
INAF--Osservatorio Astrofisico di Arcetri, Largo E. Fermi, 5, I-50125 Firenze, Italy;
\email{laura@arcetri.astro.it; galli@arcetri.astro.it}
\and
Universidad Aut\'onoma de Madrid, Departamento de F\'isica Te\'orica C-XI, 28049 Madrid, Spain
\email{eva.villaver@uam.es}
}

\date{}
\abstract
{M81 is an ideal laboratory to investigate the galactic chemical and dynamical evolution through the study of its young and old stellar populations.}
{We analyze the chemical abundances of planetary nebulae and \hii~regions in the M81 disk for insight on galactic 
evolution, and compare it with that of other galaxies, including the Milky Way.}
{We acquired Hectospec/MMT spectra of 39 PNe and 20 \hii\ regions,  with 33 spectra viable for temperature and abundance analysis.
Our PN observations represent the first PN spectra in M81 ever published, while several \hii\ region spectra have been published before,
although without a direct electron temperature determination.
We determine elemental abundances of helium, nitrogen, oxygen, neon, sulfur, and argon in PNe
and \hii~regions, and determine their averages and radial gradients.}
{The average O/H ratio of PNe compared to that of the \hii\ regions
indicates a general oxygen enrichment in M81 in the last $\sim$10 Gyr.
The PN metallicity gradient in the disk of M81 is $\Delta{\rm log(O/H)}/\Delta{\rm R_G}$=-0.055$\pm$0.02. Neon and sulfur in PNe have a radial 
distribution similar to that of oxygen, with similar gradient slopes. 
If we combine our \hii\ sample with the one in the literature we find a  possible mild evolution of the gradient slope, with 
results consistent with gradient steepening with time. Additional spectroscopy is needed to confirm this trend.
There are no Type I PNe in our M81 sample, consistently with the observation of only the brightest bins of the PNLF, the galaxy metallicity, and the
evolution of post-AGB shells.}
{Both the young and the old populations of M81 disclose shallow but detectable negative radial metallicity gradient, which could be slightly steeper 
for the young population, thus not excluding a mild gradients steepening with the time
since galaxy formation. During its evolution M81 has been producing oxygen; its total oxygen enrichment exceeds that of other nearby
galaxies.}

\authorrunning{Stanghellini, L. et al.}
\titlerunning{M81 draft}
\maketitle

\section{Introduction}

The study of chemical evolution of galaxies hinges on the confrontation between sets of evolutionary 
chemical models and arrays of observational data. Comparison of models 
with reliable abundance determinations 
are needed to understand the framework of galaxy evolution, and to test star formation 
history, stellar nucleosynthesis, and the mechanisms of galaxy assembly and formation. 

Galactic metallicity gradients, and especially their time evolution, are important probes of galaxy 
formation and evolution. Observationally, most studied galaxies display a radial metallicity gradient with a negative slope,
indicative of higher metallicity toward the inner galaxy. Time variation of gradient slopes in galaxies is
hard to come by, given the lack of reliable time-tagging metallicity probes. From the viewpoint of theory,
different models predict opposite time evolution of the metallicity 
gradient, showing the sensitivity of this physical parameter to the adopted parameterization of the 
processes (e.g., Magrini et al. 2007). A key factor to forwarding our knowledge in this field
is the acquisition of high quality spectra to obtain abundances for a variety of probes,
in a wide selection of different galaxies, including our own,
 to cover as much as possible the parameter  space of initial masses, metallicities, and galaxy type.

The planetary nebula (PN) abundances of elements that are invariant during the evolution of low- 
and intermediate-mass stars (the so-called $\alpha$-elements) probe the environment at the epoch of the 
formation of the PN progenitors. Oxygen, neon, sulfur, and argon are, for the most part, manufactured in 
high mass stars (M$>$8 M$_{\odot}$), thus their concentration in PNe across galactic 
disks probe the metallicity gradient in remote epochs. The metallicity gradients of PNe, compared 
to gradients of young stellar populations (such as those given by \hii~regions) allow the study of 
the chemical evolution of galaxies. Galactic metallicity gradients are measured in terms of 
$\Delta$log (X/H)/$\Delta{\rm R_G}$ dex kpc$^{-1}$, where X/H is the abundance of element X with respect to hydrogen and 
R$_{\rm G}$ is the galactocentric distance. 

Planetary nebula gradients in the Milky Way indicate that the 
metallicity decreases inside out by 0.01 to  0.07 dex kpc$^{-1}$ (e.~g., Maciel \& Quireza 1999, Henry et al. 
2004, Stanghellini et al. 2006, Perinotto \& Morbidelli 2006). The most recent of these studies seem to 
agree that the PN metallicity gradients are rather
flat, i.e., their slope is not steeper than $\sim$-0.02 dex kpc$^{-1}$, 
and only slightly shallower than that derived from \hii~regions, implying that the Galactic metallicity gradient does not
vary very much within the  time frame of Galactic evolution. The metallicity gradients based on Galactic PNe are hindered 
by the uncertainties in PN distances, by the interstellar absorption in the direction of the Galactic 
center, and by the paucity of PN observations in the direction of the Galactic anti-center. These 
factors, together with the interest of studying metallicity gradients of other spiral galaxies 
of different morphological types and belonging to different groups, prompt us to investigate the PN (and 
\hii~region) metallicity gradients in the M81 disk. 

M81 is the largest member of the nearest interacting group of galaxies 
at a distance of 3.63$\pm$0.34 Mpc (Freedman et al. \cite{freedman01}). 
Its two brightest companions, M82 and NGC 3077, are located within a projected distance of 60 kpc. The disturbed nature of the 
system is evident from the observations in the  21~cm emission line of \hi, showing  extended tidal streams and 
debris between the galaxies (Gottesman \& Weliachew \cite{gottesman75}; Yun et al. \cite{yun94}). 
The  simulation of Yun (\cite{yun99}) explains many of these tidal features as being the result of close encounters between M81 
and each of its neighbors $\sim$200-300 Myr ago. The most prominent debris
(e.g., Sabbi et al. 2008, Weisz et al. \cite{weisz08}) are considered tidal dwarf galaxies, 
new stellar systems that formed in gas stripped from interacting galaxies.

The $\alpha$-element abundances of PNe and \hii\
regions in M81 have remained elusive, with the exception of 17 \hii~regions (Garnett \& Shields 
1987, GS87). With the PNe (Jacoby et al. 1989, Magrini et al. 2001) and \hii~regions (Lin et al. 2003) 
locations, and their de-projected (PA=157$^0$ and inclination=59$^0$, Kong et al. 2000) galactocentric distances available 
we are in the position to push forward a determination of abundance gradients from PNe and \hii~regions 
in the M81 disk. The location of M81 within its  group 
will allow us to study the effect of galaxy interactions and mergers in the evolution of a galaxy and of its abundance gradient.  

In order to sample the M81 disk to the 
necessary S/N for abundance analysis we need spectroscopy with a 6-8m class telescope. 
MMT/Hectospec medium-resolution spectroscopy offers the unique opportunity to study 
both the PNe and the \hii~ region populations with the same setup, during the same night, and adopting the same analysis 
techniques, thus avoiding most biases that commonly affect the comparison of chemical abundances 
in population of different ages (e.g. Fe/H from old RGB stars is difficult to compare with O/H 
from the present time \hii~  regions). We present here the results from the study of a sample of bright 
PNe in M81, with the goal of obtaining the metallicity and explore their gradients within the M81 disk.
Their properties as a group are discussed, and compared with those of \hii~ regions from both our
MMT observations and from the literature. The combined properties of M81 PNe and \hii~ regions are also compared with 
their homologs in other nearby galaxies and in the Milky Way.

In $\S$2 we describe the observation and analysis techniques, and give the measured 
fluxes and calculated abundances. $\S$3 includes the analysis of the derived abundances and 
the determination of the metallicity gradients, $\S$4 presents the discussion of our results, and the conclusions are in
$\S$5.

\section{The MMT observations: data reduction and analysis}

\subsection{Observations}

We obtained the spectra of PNe and \hii~ regions in M81 with the 
MMT Hectospec fiber-fed spectrograph (Fabricant et al. 2005).  The
spectrograph was equipped with an Atmospheric Dispersion Corrector and
it was used with a single setup: 270 mm$^{-1}$ grating at a dispersion
of 1.2 \AA ~pixel$^{-1}$.  The resulting total spectral coverage
ranged from approximately 3600 \AA\ to 9100 \AA, thus including the basic emission-lines necessary
for the determination of physical and chemical properties.  

The PNe were selected among those observed 
with the INT through with \oiii, \ha, and Stromgren Y 
filters (Magrini et al. 2001), which allowed us to 
cover the whole 
field
of M81, and to identify a large number of disk PNe at large galactocentric distances.  
An earlier survey by Jacoby et al. (1989) was limited to the M81 bulge.
It is worth noting that the INT images allow us to have accurate positions ($<$0.5\arcsec) and 
finding charts for the selected PNe and \hii\ regions, a crucial information given the crowding of 
the inner and spiral arm fields of M81. 

The first target selection was performed by scaling the PN brightness in \oiii~ with
those of M33 PNe, whose medium-resolution spectra have been observed by us using the identical technique
(Magrini et al., 2009, 2010, hereafter M09, M10).
We also endeavored to have PN targets at several galactocentric distances, in order to characterize 
the PN metallicity gradient. The instrument deploys 300 fibers
over a 1 degree diameter field of view and the fiber diameter is$\sim$ 1.5\arcsec\ .
The projected size of M81 on the sky  is smaller than the whole 
field
 of view of MTT/Hectospec, and we took advantage of this by placing some of 
the
fibers 
in the outermost periphery of M81, where some \hii~ regions and few PNe, both belonging 
to the outer disk and to the intra-group population, have been identified.
In addition, since the available catalog of \hii~ regions by Lin et al. (2003) is based on \ha~ photometry, 
our INT \oiii~ observations have been useful to select \oiii-bright \hii~regions essential for chemical 
abundance determination.  As stated by Magrini et al. (2001), we expect the misclassification of \hii\ regions into PNe and vice-versa to be lower than$\sim$3$\%$ of the whole sample.

The observations were carried out in queue mode in four runs during the months of November 
and December 2008. Each of the observing runs consisted of 3 to 5 exposures of 1800s each, for a total
of 16 exposures (or 8 hours of observation).
We used the same fiber setup during all observing runs, and thus PNe and HII regions were observed within the same conditions, including the same exposure time. 
Priority was given to PNe when placing the fibers, since a 
set of \hii~region spectra already exists in the literature (GS87). A large number of fibers were devoted to sky measurements, to ensure that sky spectra were obtained in several positions around each target.

Cosmic rays were removed with the appropriate {\it hectospec.hcosmic} routine in the hectospec analysis package. Cosmic rays are detected via subtraction of multiple images, flagging the high or low pixels. The flagged pixels are  interpolated over, and the resultant images are combined by average.  This method  avoids problems with clipping that other programs suffer from. In order to perform the sky subtraction we first inspected the sky spectra and eliminated those with unusual signals. Then we averaged the six spectra closer (in fiber location) to the targets, scaled the result
to match the object spectrum, and subtracted.

The spectral calibration was achieved with the standard star Hilt600 (Massey et al.~\cite{massey88}), observed
during the first night. This star was chosen by its excellent spectral coverage compared to the 
spectra of the PNe. It is worth noting that our analysis does not hinge on flux calibration, rather spectral calibration
and flux ratios, thus this approach is sufficient to the science goals. Other stars were observed during the run nights, 
but the excellent spectral coverage of Hilt600 makes it the ideal choice for this type of calibration. Variations in sky
conditions are taken into account since sky subtraction was performed before flux calibration.

In Table 1 we give the IDs, equatorial coordinates, and equivalent \oiii~
magnitude of the planetary nebulae and \hii~regions observed with MMT medium-resolution
spectroscopy. All PN IDs are from Magrini et al. (2001), except for PN4 and PN5, which are identified PNe with
unpublished positions  (Magrini, private communication). 
There are 39 PNe  and  20 \hii~regions whose spectra have been acquired and 
analyzed. With the exception of PN~49m and PN~79m, whose spectral
lines are below the detection limit, we detect at least the major spectral lines in each target.

\begin{table}
\caption{Observing log}
\label{table_1}
\begin{tabular}{llll}
\hline\hline
Id.&$\alpha_{\rm J2000}$&$\delta_{\rm J2000}$&  m$_{\lambda5007}$\\
(1)& (2)& (3)& (4)\\
\hline
PN~3m&9:54:26.54&69:10:30.5&	24.41\\
PN~4&9:54:47.50&69:35:42.0&	24.15\\
PN~4m&9:54:27.66&69:13:55.0&	24.45\\
PN~5&9:54:53.80&69:08:01.0&	23.87\\
PN~6m&9:54:36.70&69:17:00.5&	24.39\\
PN~7m&9:54:38.52& 69:10:36.8&       24.53\\
PN~8m& 9:54:45.50&69:08:06.6&       24.06\\
PN~9m&9:54:46.52&69:05:39.0&	23.66\\
PN~11m&9:54:53.15&69:13:32.2&	24.51\\
PN~14m&9:54:09.28&69:11:31.0&	24.75\\
PN~15m&9:55:01.01&69:12:04.3&	24.45\\
PN~17m&9:55:10.09&69:02:37.1&	24.24\\
PN~28m&9:55:13.62&69:07:10.3&	23.91\\
PN~29m&9:55:14.78&68:55:30.0&	23.56\\
PN~32m&9:55:15.70&69:08:39.8&	24.23\\
PN~33m&9:55:15.96&69:13:31.7&	24.30\\
PN~38m&9:55:19.47&68:54:45.7&	24.39\\
PN~41m&9:55:02.58&69:06:26.3&	23.77\\
PN~45m&9:55:21.81&69:07:50.6&	23.86\\
PN~49m&9:55:23.38&69:02:50.9&	23.66\\
PN~63m&9:55:28.25&69:01:26.5&	24.26\\
PN~70m&9:55:30.62&69:07:56.3&	23.66\\
PN~79m& 9:55:34.58& 69:09:00.9&       24.49\\
PN~93m&9:55:39.07&69:05:42.1&	23.48\\
PN~117m&9:55:50.11&68:59:48.4&		24.04\\
PN~121m&9:55:51.46&69:05:14.9&		23.56\\
PN~127m&9:55:52.93&69:10:52.1&		24.60\\
PN~128m&9:55:53.02&69:06:15.2&		24.19\\
PN~136m&9:55:06.11&68:57:18.5&		24.60\\
PN~137m&9:55:07.29&69:12:10.4&		24.47\\
PN~147m&9:56:20.74&68:58:26.8&		24.41\\
PN~149m&9:56:27.09&69:05:26.4&		23.66\\
PN~153m&9:56:28.70&69:00:32.8&		23.56\\
PN~157m&9:56:31.99&69:03:31.3&		24.09\\
PN~158m&9:56:32.84&69:03:58.8&		24.23\\
PN~160m&9:56:04.48&69:06:36.0&		24.31\\
PN~166m&9:56:08.41&69:08:37.8&		24.49\\
PN~167m&9:56:08.54&69:03:56.0&		24.53\\
PN~169m&9:56:09.87&68:59:56.0&		24.39\\
\hline
\hii~4     &9:54:34.67 &69:05:52.2&\\
\hii~5     &9:54:36.1   &69:06:30.5&\\
\hii~31 &9:54:43.62 &69:03:29.9&\\
\hii~ 42&   9:54:47.66 &69:02:18.7&\\ 
\hii~72 &9:54:52.93 &69:08:51.8&\\
\hii~78& 9:54:53.84& 69:01:08.8&\\
\hii~79 &9:54:54.28 &69:10:56.4&\\
\hii~81 &9:54:54.53 &69:09:30.5&\\
\hii~121& 9:55:02.87 &69:04:49.3&\\
\hii~123 &9:55:03.64 &69:10:54.6&\\
\hii~133 &9:55:04.98 &69:09:53.6&\\
\hii~201 &9:55:22.82 &69:09:52.9&\\
\hii~213& 9:55:26.01& 69:11:58.9&\\
\hii~228 &9:55:32.66 &69:11:55.4&\\
\hii~233 &9:55:34.59 &69:08:31.1&\\
\hii~249&   9:55:37.61 &69:12:03.6&\\
\hii~262 &9:55:41.42 &69:11:12.2&\\
\hii~282&   9:55:46.50 &68:59:24.6   &\\
\hii~325 &9:55:58.48 &69:09:9.9&\\
\hii~328&   9:55:59.29 &68:59:37.6 &\\
\hii~348 &  9:56:03.23 &69:07:38.3  &\\
\hii~352  & 9:56:03.79 &69:09:37.8 &\\
\hii~384 &  9:56:08.33 &69:05:14.8 &\\
\hii~403 &9:56:11.78 &69:07:44&\\

\hline
\end{tabular}
\end{table}

In Table 2, available online in its completeness, we give for each target the ion and wavelength (columns 1 and 2) of the observed emission line, then the 
relative observed flux
(column 3), its uncertainty (column 4), and the 
line intensity (column 5) obtained with the extinction correction given as a header for each target. All fluxes and intensities are normalized for F$_{\rm H\beta}$=100 and 
I$_{\rm H\beta}$=100. Both PNe and \hii~regions are listed in Table 2. 
Target names are given for each target at the head of
the corresponding flux list.
The emission-line fluxes were measured with the
package SPLOT of IRAF\footnote{IRAF is distributed by the National
Optical Astronomy Observatory, which is operated by the Association of
Universities for Research in Astronomy (AURA) under cooperative
agreement with the National Science Foundation}.  Errors in the fluxes
were calculated taking into account the statistical error in the
measurement of the fluxes, as well as systematic errors of the flux
calibrations, background determination, and sky subtraction. Observations of same
objects during different runs were stacked together and the resulting spectra have
been used for the analysis.

\begin{table}
\caption{Observed Fluxes and Line Intensities} 
\label{table_2}
\begin{tabular}{crrrr}
\hline\hline

Ion      & $\lambda$(\AA) &$F_{\lambda}$     &  $\Delta(F_{\lambda})$  &     $I_{\lambda}$ \\      
(1)&(2)&(3)&(4)&(5)\\
\hline
&&&&\\
PN~3m, c$_{\beta}$=0.502$\pm$0.027\\
&&&&\\

\oii    &  3727   & 373.9    &   58.6   &       517.3     \\
\hg     &  4340   & 50.0   &   3.5 &       58.0   \\
\hb       &  4861   & 100.0    &   4.5  &       100.0    \\
\oiii   &  4959   & 35.4    &   3.8   &   	34.4  \\    
\oiii   &  5007   & 104.2    &   5.2   &   	99.9    \\  
\nii    &  5755   & 1.3    &   0.3   &   	1.1      \\
\hei      &  5876   & 9.1    &   0.3   &   	7.2      \\
\nii    &  6548   & 59.8    &   3.5   &   	42.2      \\
\ha       &  6563   & 412.0   &   6.9 &       290.1   \\
\nii    &  6584   & 142.5    &   4.5   &   	100.0  \\    
\sii    &  6717   & 60.3    &   3.8   &   	41.4      \\
\sii    &  6731   & 44.3    &   3.4   &   	30.4      \\
\hline
\end{tabular}

\end{table}

We derived \cbeta~ from the \ha/\hb\ flux ratios, as described in M09, and used the Galactic extinction curve by Mathis (1990)
to correct the observed flux for interstellar extinction. 
In a few PNe the derived intensity of the \hg\ line is more than 30$\%$ off the theoretical prediction (I$_{4340}$=0.466 for T$_{\rm e}$=10,000 K,
case B, Oserbrock \& Ferland 2006). 
The reason for this offset could be an uncertain \cbeta. The average \cbeta\ calculated for our PNe is 0.7$\pm$0.3, a reasonable value
if we consider the average \cbeta\ of Galactic PNe (from Cahn et al. 1992, $<$\cbeta$>$=0.95) and  the relative metallicities of the Galaxy to M81.
We will recall these outliers later in the discussion and indicate them with different symbols in the Figures. 

To our knowledge, this is the first attempt in the literature to detect PN emission line fluxes in M81 
thus we cannot extend our
discussion to comparison with previous work. The spectral quality is excellent. 
In Figure 1 we show a few examples of our M81 PN spectra, where the 
plasma and abundance diagnostic lines are well evident, and the S/N obtained with our 
observations makes us confident on the data quality.

Several \hii~regions have been already observed by GS87, where the Authors refer to the catalog
by Hodge \& Kennicutt (1983) for coordinates. Since the latter are given in image coordinates rather than 
Galactic coordinates, it has been hard to eliminate from our sample of \hii~regions those that have been already 
analyzed by GS87 or comparing targets from the two sets. But by comparing the position visually,
we did not find any obvious overlap between the two samples, thus cannot compare the results.

\begin{figure*} 
\centering 
\includegraphics[width=17cm]{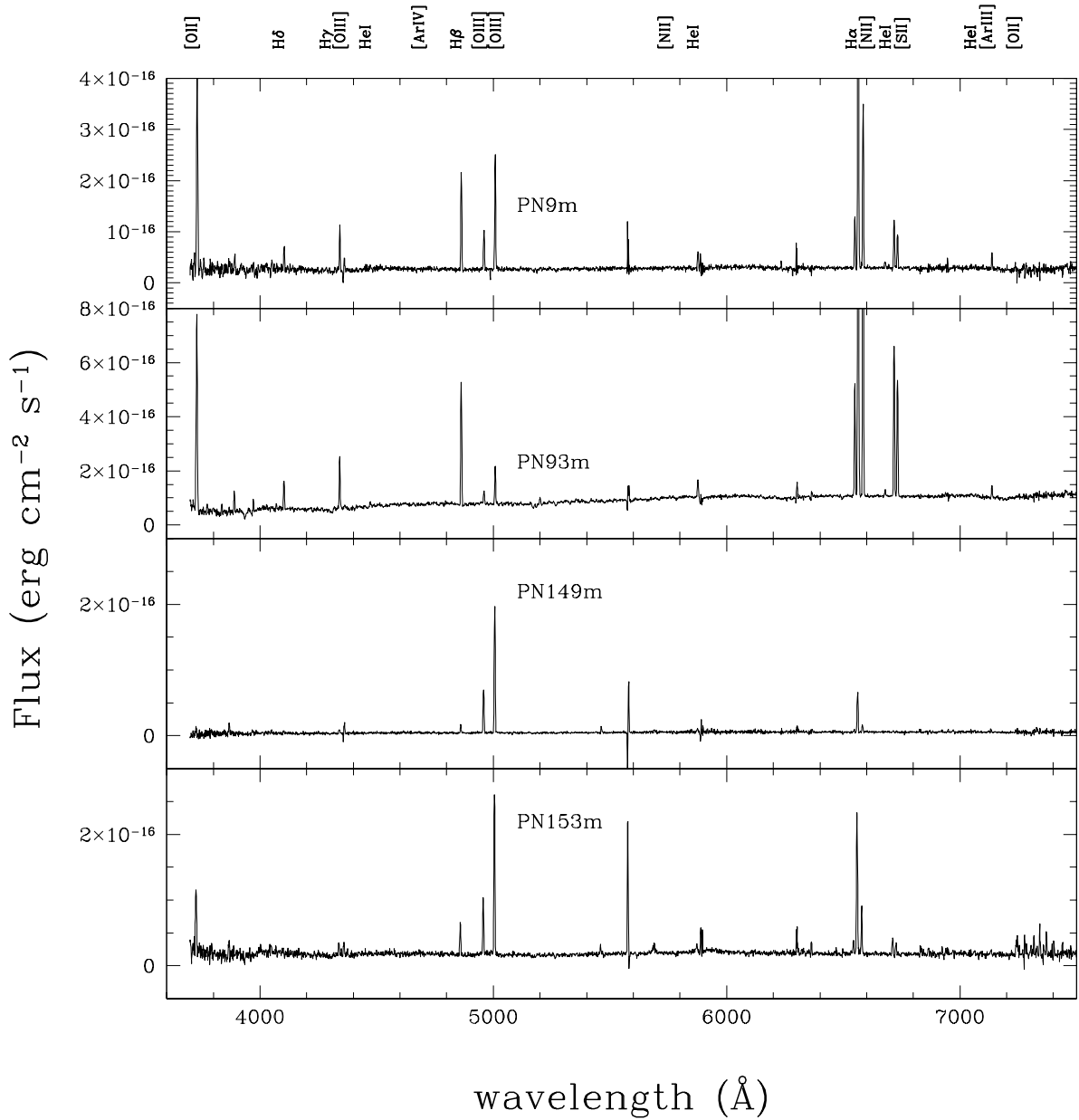} 
\caption{Sample of medium-resolution spectra of M81 PNe.} 
\label{} 
\end{figure*}

\subsection{Plasma diagnostics}

We perform plasma diagnostics by using the 5-level atom model included in the {\it nebular}
analysis package in IRAF/STSDAS (Shaw \& Dufour~1994), consistently with our work in M33 (M09, M10). 
To calculate densities we use the doublet of the sulfur lines
\sii$\lambda\lambda$6716,6731, while for the electron temperatures we 
use the ratios \oiii$\lambda$4363/($\lambda$5007 +$\lambda$4959),  \nii$\lambda$5755/($\lambda$6548 +
$\lambda$6584), and \siii$\lambda$6312/($\lambda$9059 +
$\lambda$9532).   In some cases, even if the sulfur lines are available, 
their ratio is outside the diagnostic ranges for density calculation (e.g., Stanghellini \& Kaler 1989), and we
assumed a density of 10$^3$cm$^{-3}$. 
We calculate the medium- and low-excitation temperatures
respectively from the \oiii\ and \siii, and from the \nii\ line ratios, respectively (see also
Osterbrock \& Ferland 2006,$\S$5.2).

\subsection{Abundance analysis}

Of the 39 PNe with emission line spectra,  \oiii\ and \nii\ temperature diagnostics are available respectively
for 10 and 12 PNe, with only 3 PNe in common. In the analysis of the M33 PNe (M09) we were able to
infer the electron temperature for targets whose plasma diagnostics were not available by 
using the correlation between the electron temperature and the \heii\ line intensity. In the case of M81 none of the PNe show the \heii\ emission;
as a consequence, abundance analysis was possible only for the 19 
PNe where electron temperature was diagnosed directly. 
There are also 14 \hii\ regions whose temperature analysis was based on auroral lines, and their plasma and abundance analysis was performed in a similar way than for the PNe.
Electron temperature diagnostics for more than one excitation state was available for 5 of the 14 regions.

The plasma diagnostics are given in Table 3, available online, together with the ionic abundances of the PNe and \hii\ regions. Column (1) gives the diagnostics, 
and column (2) the value as determined from our analysis. The targets are listed sequentially, as in Table 1, and the first line for each target 
gives the target name.

\begin{table}
\caption{Plasma Analysis and Abundances}           

\label{table_3}
\begin{tabular}{ll}

\hline\hline
(1)& (2)\\
\hline

	                Name &    PN~3m\\
		                 \te\nii           &    8893           \\
                          N$_{\rm e}$\sii         &   100  \\
                          \hei/H           &       0.051           \\
                                 He/H           &       0.051           \\
                                \oii/H           &   3.320$\times10^{-4}$         \\
                               \oiii/H           &   5.440$\times10^{-5}$         \\
                               ICF(O)           &       1.0           \\
                                  O/H           &   3.864$\times10^{-4}$         \\
                          12+log(O/H)           &   8.587             \\
                                \nii/H           &   2.916$\times10^{-5}$         \\
                               ICF(N)           &       1.164           \\
                                  N/H           &   3.394$\times10^{-5}$         \\
                          12+log(N/H)           &   7.531             \\
                                \sii/H           &   2.250$\times10^{-6}$         \\
                               ICF(S)           &       1.001           \\
                                  S/H           &   7.066$\times10^{-6}$         \\
                         12+log(S/H)           &   6.849             \\
 
  \end{tabular}

\end{table}

The abundance analysis of the PNe and the \hii\ regions was performed following the prescription given by M09, M10; 
abundances of all elements except helium where 
calculated with the ionization correction factors (ICFs) given in Kingsburgh \& Barlow (1994) for the case were only
optical lines are detected. Where both \teoiii\ and \tenii\ were available we used \teoiii\ to derive the abundances of medium and high excitation
ions, and \tenii\ for the low excitation ones. In several \hii\ regions the temperature from the auroral line of [SIII]  were also available, corresponding 
to transition of intermediate excitation. If only one of the electron temperature diagnostics was available we
use that temperature to derive abundances for all ions. Helium abundances were calculated with the formulation of
Benjamin et al. (1999) in two density regimes, and taking into account Clegg's (1987) collisional populations, as in M09.
The sulfur abundances for our PNe are mostly based on \sii\ lines, thus highly unreliable. We used the interpolation to the Galactic 
PN data by Kingsburgh \& Barlow to infer \siii/\sii\ from the \oiii/\oii\ ratios, where possible, and obtained an estimate of \siii, which is observed only in three PNe. 
In order to check whether the sulfur correction is sensible, in Figure 2 we plot log${\rm (S^{2+}/S^+)}$ versus log${\rm (O^{2+}/O^+)}$ for the Kingsburgh \& Barlow 
sample. On the same plot we locate our M81 PNe with measured \siii.
The solid line, representing ${\rm S^{2+}/S^+=4.677\times log(O^{2+}/O^+)^{0.433}}$, 
has the correct slope for both the Galactic and M81 data, but it overcorrect slightly for \siii. Since the \siii\ lines we observe in M81 are quite faint, the 
relation by Kingsburgh \& Barlow (1994) seems reasonable.

\begin{figure} 
\resizebox{\hsize}{!}{\includegraphics{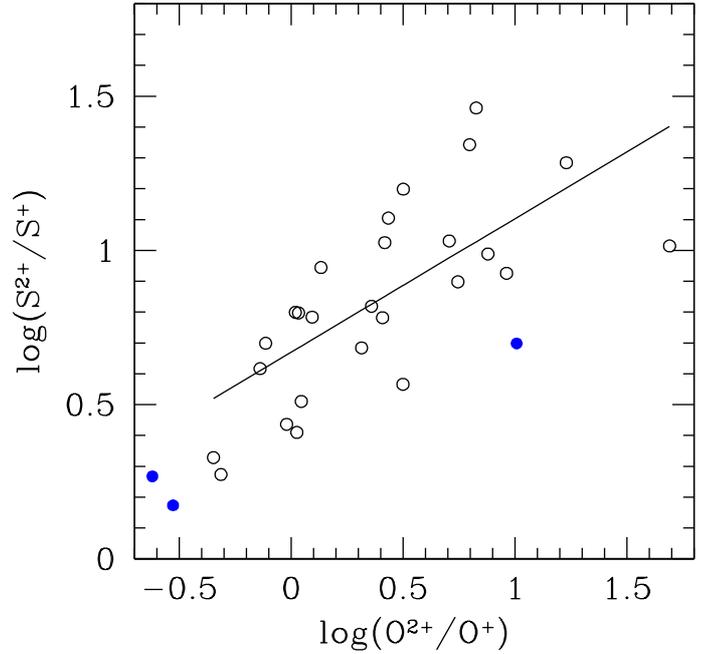}} 
\caption{The log(S$^{2+}$/log(S$^+$) versus log(O$^{2+}$/log(O$^+$) correlation (solid line), based on the
Galactic PNe from Kingsburgh \& Barlow (1994, open circles). Our targets are shown with filled circles.} 
\label{} 
\end{figure}

The formal errors in the ionic and total abundances were 
computed taking into account the uncertainties in (i) the observed 
fluxes, 
(ii) the electron temperatures and densities, and (iii)  \cbeta.
We performed complete formal error propagation for a subsample of target, representing  a spread of flux error and magnitudes,
and realized that the errors in the electron temperature
completely dominate the errors in the ionic abundances for all elements except helium. In particular, the flux uncertainty of the weaker of the diagnostic lines
in the electron temperatures, $\lambda$5755 for \tenii\ and $\lambda$4363 for \teoiii,  determine the errors in the abundances. 
In Table 4 we give the error analysis for the PN subsample, including the ones with the minimum and maximum flux errors; column (1) gives the PN name,
column (2) gives the relative flux error for the weaker line in the temperature diagnostics (in PN~9m we could calculate both \tenii\ and 
\teoiii, thus we give the two errors), columns (3) and (4) give the relative errors in \tenii\ and \teoiii\ respectively, and finally columns (5) through (7)
give the abundance errors, in dex. 
We derived the errors for all other targets by interpolation of the $\Delta$F$_{\lambda}$-$\Delta$A(X) curves. 
In the case of the helium abundances we run the error analysis by using the formulae by Benjamin et al. (1999), and found that the 
factor determining the uncertainties is the relative error on the helium flux lines. 
The abundances, their uncertainties, together with the galactocentric distances (see $\S$3.1) are given in Table 5.  
 
It is worth noting that another source of error that we could not estimate explicitly is intrinsic to the ICF method. 
The uncertainties due to the ICF are very moderate for oxygen, but can be quite large for the other atoms, especially sulfur,
thus these abundance uncertainties should be considered as formal errors, but might underestimate the real uncertainties.
Another source of  abundance uncertainties is the use of a single electron temperature for both low- and high-ionization
ions. This regards the 16 PNe and  9 \hii\ regions where only one choice of the electron temperature 
was available.  We have analyzed the sample of targets where two electron temperatures were available, and found that the 
average difference between electron temperatures of the order of 20$\%$. We tested that the resulting 
atomic abundance uncertainties are lower than those due to flux errors on the auroral lines. 

The sets where we have oxygen and neon in common is limited to 6 PNe and 7 \hii\ regions.
There is a clear correlation between these two $\alpha$-elements in both sets, with correlation coefficients R$_{\rm xy}$=0.7 and 0.9
respectively for PNe and \hii\ regions. This high correlation is expected, since 
both elements are produced in the Type II SN environment. The large uncertainties in the sulfur abundances make the comparison between 
oxygen and sulfur less meaningful, although the two sets have higher overlap than the previously described pair (16 PNe and 14 \hii\ regions in common), 
the correlation coefficient is only 0.4 and 0.3 for the two sets.

\begin{figure} 
\resizebox{\hsize}{!}{\includegraphics{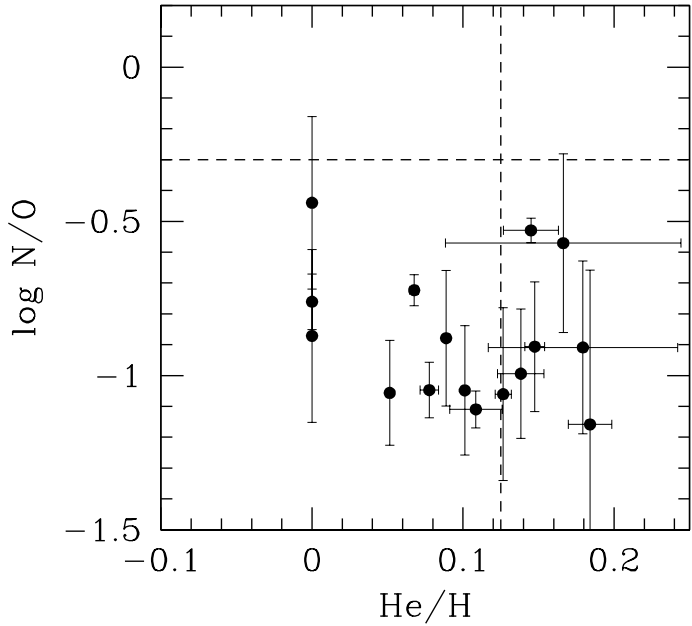}} 
\caption{log (N/O) vs. He/H; broken lines show He/H=.125 and log(N/O)=-0.3. Type I PNe would be located in the 
upper-right quadrant.} 
\label{} 
\end{figure}

Figure 3 shows the log(N/O) vs. helium plot,  to classify Type~I PNe. 
These are the nitrogen and helium-enriched PNe, whose progenitors have likely undergo the third dredge-up
and hot bottom-burning, and thus are likely to have higher progenitor masses
 (Peimbert \& Torres-Peimbert 1983, Marigo 2001). The top right portion of the plot, where
He/H$>$0.125 and log(N/O)$>$-0.3, is the locus of Type I PNe as determined for the Milky Way by Perinotto et al. (2004). Since the metallicity of M81 
is similar to that of the Milky Way, we can use the same diagnostic plot for M81 PNe.
Clearly, there are no Type I PNe among those observed by us in M81. In the next section we discuss the population of the
observed PNe and its implications for gradients and evolutionary studies, and show that indeed we do not expect an observed
population of Type I PNe among our targets.

In Table 5, at the bottom of each set of targets, we give the average abundances, relative to hydrogen (linear for helium, and in logarithmic format 
for the other elements. Uncertainties represent ranges of abundances).
The average oxygen abundance of M81 PNe (2.83$\pm$2.01$\times$10$^{-4}$) is 
lower that what has been inferred for the Galactic disk PNe (4.49$\pm$2.28$\times10^{-4}$ for Type II PNe,
Stanghellini \& Haywood 2010) and about 
1.6 higher than the average oxygen abundance of Type II  I PNe in M33 (M09).

The \hii~regions observed by us span a small  fraction of the galaxy radius (5.8 to 9.8 kpc from galaxy center).  
The average O/H value for these regions, (3.42$\pm$1.78)$\times$10$^{-4}$, should be compared to the oxygen abundance
of PNe in the same radial range, which is (1.96$\pm$1.27)$\times$10$^{-4}$, showing a 0.25 dex oxygen enrichment in this area of the 
galaxy. The comparison between PNe and \hii\ region average abundances  of other elements is very limited, since the sets have 2 or 3 objects in common 
in the 5.8-9.8 kpc region.

\subsection{The population of M81 PNe.}

The PNe whose medium-resolution spectra are presented here are selected form the bright end of the planetary nebula luminosity function (PNLF)
of the M81 disk. They have two peculiarities, as a group: first, we do not detect the \heii\ $\lambda$4686 emission lines 
in any of them; this is a signature of moderately low central star temperature (less than $\sim$100,000 K). Second, none of the PNe whose 
abundance analysis was performed (i.e., whose electron temperature has been measured from the emission lines)
is of Type I, thus they must be the progeny of stars with turnoff masses (M$_{\rm to}$)
in the$\sim$1-2 M$_{\odot}$ range. 

The properties of the M81 PN sample 
are important to define their nature when we compare the population of M81 PNe to those of other galaxies.
M81 has similar metallicity to that of the Milky Way (Davidge 2009). In M81, as in most external galaxies, 
the observed PN population belong to the brightest few magnitude bins. The 
nature of the brightest PNe varies notably  with the metallicity of the host galaxy. In their runaway evolution from the AGB, the PN central stars (CS)
reach hot temperature while keeping very bright; later, both brightness and temperature decline.  A PN is visible only when 
the CS temperature is high enough to ionize hydrogen, and the nebula is optically thin to the \ha\ radiation (e.g., Stanghellini \& Renzini 2000). 
At relatively high metallicities such as those of M81 or the Galaxy,
the brightest PNe are not those with the hottest CSs (e.g., Villaver et al. 2002). In fact, the post-AGB shells with very hot CSs are statistically still enshrouded 
in dust at early phases or their evolution, thus still thick to the \ha\ radiation. A lower metallicity implies a shorter thinning time 
(K\"aufl et al. 1993). Plenty of Type I PNe have been observed within the bright end of  the PNLF of low metallicity galaxies such as
 M33, the LMC, and the SMC; this is not the case in M81, where the high metallicity prevents massive progenitor PNe (i.e., the Type I) to populate the
 bright end of the PNLF.

To illustrate this point, 
in Figure 4 we show the luminosity function (PNLF) of  Galactic PNe
in the \hb-equivalent absolute magnitude.  The \hb\ fluxes are from Cahn et al. (1992), and the PN distances from 
Stanghellini et al. 2008 (see also Stanghellini \& Haywood 2010). The solid histogram represents the homogeneous
Galactic sample of 331 PNe where both distance and \hb\ flux are known, and where spectroscopy has been performed and
a lower limit or a detection of the \heii\ $\lambda$4686 line has been acquired. The shaded histogram is the subsample
of Galactic PNe with I$_{4686}$/I$_{\beta} <$ 5$\%$. 
It is clear that at least 80-90$\%$ of the bright Galactic PNe 
do not show the \heii\ lines, in qualitative agreement with not finding this emission line in the M81 PN sample.

\begin{figure} 
\resizebox{\hsize}{!}{\includegraphics{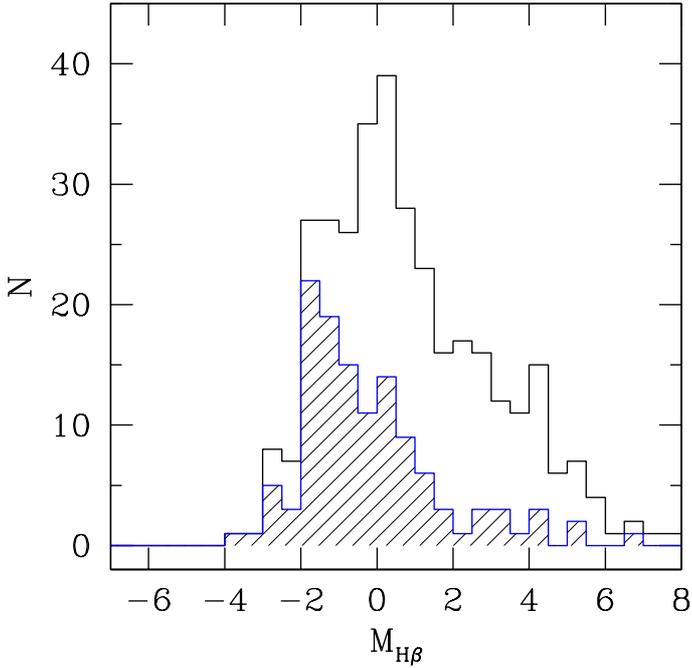}} 
\caption{The Galactic PNLF in the light of \hb\ for a homogeneous sample of PN (unshaded histogram) and for
a subsample of PNe with no \heii\ emission lines (shaded histogram, see text). Data are from Cahn et al. 1992, Stanghellini \& Haywood 2010.} 
\label{} 
\end{figure} 

In order to quantify the expectation of detecting the  \heii\ emission line in a  given PN population of a certain metallicity
we use K\"aufl et al.'s (1993) models to derive PN opacity, where the non-evolving term of the opacity,$\tau_{\lambda}(0)$, scales directly with 
the dust-to-gas ratio (see their Eq. 10). We also express this ratio as expected at different metallicities,
M$_{\rm d}$/M$_{\rm g}\sim0.01$(O/H)/(O/H)$_{\rm MW}$ (Draine 2009), where the right hand side term refers to
the ratio of the oxygen vs. hydrogen abundance in a given galaxy compared to that of the Milky Way. 
To obtain the ratio for M81 scaled to that of the Galaxy  we use the \hii\ region oxygen abundances in M81 (4.99$\times$10$^{-4}$)
and in the Galaxy (5.25$\times10^{-4}$, Deharveng et al. 2000) to find ${\rm M_d/M_g \sim 0.01}$,
thus for a given type of M81 PN the thinning time would be similar to that of the Galactic  homologous.
We conclude that the lack of \heii-emitting PNe in M81 is totally expected, given the Galactic PNLF for both \heii--emitting and non-emitting PNe (Figure 4). 

It is interesting to compare the PN population of M81 with that of M33 as well. Magrini et al. (2010) obtained an average oxygen abundance of 
$<$O/H$>$=2.04$\times 10^{-4}$for \hii\ regions in the M33 disk, corresponding to a dust-to-mass ratio of$\sim$4$\times$10$^{-3}$, 
or, less than half that of M81. By running the K\"aufl et al.'s models for M33 we obtain (Balmer line) thinning times$\sim$1.5 times shorter
than in M81 or the Galaxy, which can account for the relatively higher frequency of \heii-emitting, bright PNe that where observed in M33 
by M09. 

We know that the brightest PNe in a given population excludes the progeny of the 
massive AGB stars, i.e., those with M$_{\rm to}>2$ M$_{\odot}$ (Stanghellini \& Renzini 2000). The sample of M81 PNe under study here
thus does not include the progeny of  the more massive AGB stars. This agrees very well with the observational fact that we do not
observe any Type I PNe in our M81 sample.

\section{Radial metallicity gradients in M81}

\subsection{The planetary nebula gradients}

The metallicity gradient of a disk galaxy is generally intended as the metallicity variation between the galactic center and
the periphery, measured radially. These gradients are essential constraints for the chemical evolution
models, and, depending on
the probe, can constrain the galactic composition at different times in evolution. Planetary nebulae with low-mass progenitors, such as
non-Type I PNe, probe the early stages of disk metallicity. We could derive the metallicity gradients in M81 from oxygen abundances, thus 
probing the $\alpha$-element evolution through the galactic disk.
We estimate the distances of our target PNe from the center of M81 by using a rotation angle of 157$^o$ and an inclination of
59$^o$. The error in the target coordinates is very low, and the major source of error in the distance calculation is the 
uncertainty in the galaxy inclination, which we take as in Connolly et al. (1972), $\pm2^o$, and we propagate to estimate the galactocentric distance uncertainties.

\begin{figure} 
\resizebox{\hsize}{!}{\includegraphics{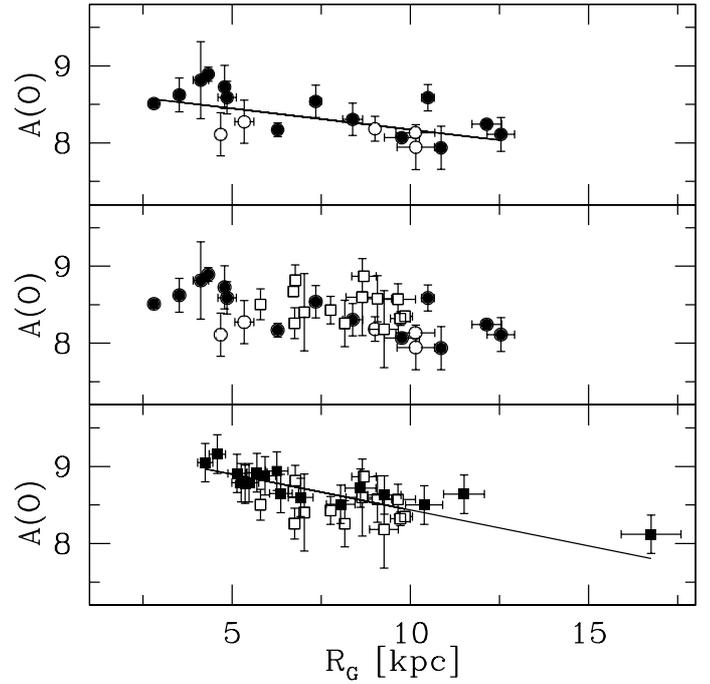}} 
\caption{Oxygen abundances vs. galactocentric 
distances for M81 targets. Top panel: the observed sample of PNe, with formal error bars and least square fit represented by solid line.
Open symbols represent those PNe whose extinction constant might be overestimated.
Middle panel: The \hii\ region sample from MMT spectra (open squares) superimposed to the PN population of the top panel.
Bottom panel: The \hii\ regions from  GS87 (filled squares) superimposed to the MMT \hii\ regions of the middle panel (open squares). The
line represents the least square fit for the combined sample, as given in Table 6. } 
\label{} 
\end{figure}

In Table 6, columns (3) and (4),  we give the metallicity gradients calculated for oxygen, neon, and sulfur. Slopes are in dex kpc$^{-1}$, while
intercepts are in dex, with the usual notation of A(X)=log(X/H)+12.
In Figure 5 (top panel) we show the oxygen gradient from the M81 PNe, where the line represent the linear fit. Only PNe are plotted in this top panel, with 
filled symbols representing the general PN population, and the open symbols indicating those PNe whose extinction correction from the \ha/\hb\ ratio does not well reproduce the line 
intensity expected for the \hg\ emission line. These open data points might thus have larger uncertainties than inferred from the formal 
error analysis. 

We perform the fits  with the routine {\it fitexy} (see Numerical Recipes, Press et al. 1992) and
obtain $\Delta{\rm log(O/H)}/\Delta{\rm R_G}$=-0.055$\pm$0.02 dex kpc$^{-1}$.  Should we exclude the open symbols in the fit we would obtain
a very similar result ($\Delta{\rm log(O/H)}/\Delta{\rm R_G}$=-0.053$\pm$0.02 dex kpc$^{-1}$).

In Figure 6 we show the metallicity gradients for sulfur (top) and neon (bottom), where the PNe are plotted with filled symbols. 
The solid lines represent the fits obtained with the  {\it fitexy} routines that consider the formal errors both in abundances and distances (see Table 6).
Fits obtained with the least square method are very similar.

\subsection{\hii~regions in the A(X)--R$_{\rm G}$ plots}

The bottom panel of Figure 5 shows the \hii~region analysis of oxygen abundances from this paper (open symbols) and from GS87 (filled symbols). Formal errorbars have 
been applied to the data obtained in this paper, while in the case of the 
GS87 sample we assume a 0.05$\%$ relative uncertainty in the distances, and a 0.25 dex uncertainty in the oxygen abundances.
It is worth noting that we have recalculated the galactocentric distances for the GS87 targets by assuming the same distance to M81 than we use  thorough this paper 
(d=3.63 Mpc) for homogeneity, although this does not affect the resulting gradient significantly. 
The composite data sample has been fitted including the distance and abundances uncertainties with the {\it fitexy} routine, to give a gradient 
($\Delta{\rm log(O/H)}/\Delta{\rm R_G}$=-0.093$\pm$0.02 dex kpc$^{-1}$) not too dissimilar, yet steeper, from that of PNe. 
A direct least square fit would give a slightly milder gradient ($\Delta{\rm log(O/H)}/\Delta{\rm R_G}$=-0.07 dex kpc$^{-1}$).
It is worth noting that both the average O/H abundance in \hii~regions and the intercept of the 
linear metallicity gradient are consistently higher that the respective values for PNe, as shown in Table 6, with PNe under abundant with respect to \hii~  regions by almost a factor of 2, on average. This is well seen in the middle panel of Figure 5, where PNe and \hii\ regions from MMT observations are plotted together.

\begin{figure} 
\resizebox{\hsize}{!}{\includegraphics{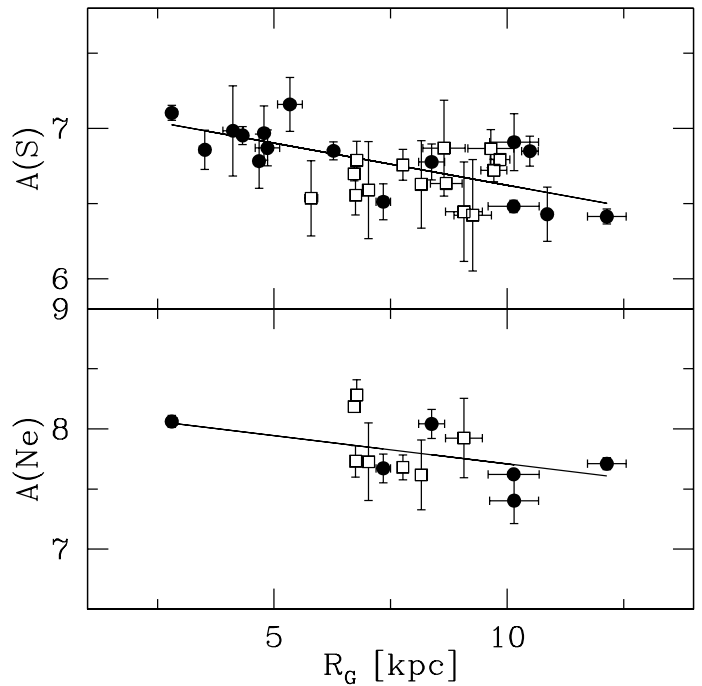}} 
\caption{Sulfur  (top panel) and neon (bottom panel) abundances from MMT observations vs. galactocentric 
distances for M81 PNe (filled circles) and \hii\ regions (open squares). The lines represent the linear fits to the PNe as discussed in the text. } 
\label{} 
\end{figure}

In Figure 6 we show the sulfur (top) and neon (bottom) abundances in \hii\ regions versus their galactocentric distances (open symbols),
plotted together with the PN data (filled symbols). Since the  \hii~regions observed with the MMT span a small fraction of the galaxy radius, 
metallicity gradients derived form these species would not be very meaningful. Within the distance ranges populated by our \hii\ region data 
we find${\rm <Ne/H>_{\hii}/<Ne/H>_{\rm PN}}$=1.1,
and ${\rm <S/H>_{\hii}/<S/H>_{\rm PN}}$=0.9.
Any further  comparison is not sensible given the limited data at hand, and we prefer to postpone a complete discussion of the sulfur and neon abundances 
in PNe as compared to the  \hii~regions of M81 when more \hii~ region spectral data with diagnostic lines relative to these atoms will be available.

\setcounter{table}{5}
\begin{table}
\caption{Metallicity Gradients}
\begin{tabular}{lrrr}
\hline\hline
X& N$_{\rm PNe}$&$\Delta$log(X/H)/$\Delta$R$_{\rm G}$& A(X)$_0$\\
(1)& (2)& (3)& (4)\\
\hline

O&	19&   -0.055$\pm$0.02& 8.72$\pm$0.18 \\
Ne&	6&	   -0.047$\pm$0.03& 8.18$\pm$0.27 \\
S&	16&     -0.069$\pm$0.01& 7.27$\pm$0.10\\

\hline
\hline
X& N$_{\rm \hii~~reg.}$&$\Delta$log(X/H)/$\Delta$R$_{\rm G}$& A(X)$_0$\\
O&  31&   -0.093$\pm$0.02& 9.37$\pm$0.24\\
\hline
\end{tabular}
\end{table}

\section{Discussion}

\subsection{Galaxy evolution through oxygen abundances}

The O/H abundances are among the most reliable in emission-line targets when only optical spectra are available. 
In the case of M81, PNe and \hii~ regions provide a meaningful comparison between young and old stellar objects, since the techniques 
of data acquisition, analysis, and abundance calculation are the same for both classes of objects. In particular, for several \hii~ regions 
presented here, those observed with the MMT, the actual data acquisition is identical that that of the PNe, thus the comparison among 
data set is even more compelling. 

It is worth emphasizing that the  \hii\ region gradient  is not based exclusively on our MMT data, which cover 
a limited radial range, but also on the results by GS87. 
The spectroscopic observations of GS87 did not allow a direct measurement of the electron temperature. 
They derived \te\  by applying an indirect calibration made through the photoionization models by Pagel et al. (\cite{pagel79}). 
The errors on the electron temperature are estimated of the order of $\pm$1200 K, mainly due to the scatter 
of the measurements on which Pagel et al. (\cite{pagel79}) based their calibrations. 
We have estimated, using the IRAF {\it ionic} task,  that an uncertainty  of $\pm$1200 K in \te\ translates 
in $\pm$0.2 dex in the ionic abundances. 
Thus, the chemical abundances of GS87 result less accurate than ours, and may introduce 
biases in the determination of the gradient.  We conservatively have assumed uncertainties of 0.25 dex for the oxygen abundances
from GS87.

As shown in Figure 5, the radial oxygen gradient slope of \hii\ regions is mildly steeper than that of PNe, with differences compatible with the 
slope uncertainties.  Since there are no Type I 
PNe in our sample it is sensible to assume that the PN progenitor (turnoff) masses are smaller than $\sim$2 M$_{\odot}$, implying a 
population older than t $\sim$1 Gyr (Maraston 1998). The lower limit for PN progenitor mass is around 1 M$_{\odot}$, corresponding
to a galactic age of t $\sim$10 Gyr. 

In Figure 7 we plot the slope of the oxygen radial gradients in M81 (squares) for PNe and \hii~ regions, against  the mean age of the 
population considered. \hii~ regions are placed at t=0, while PN ages are inferred from the presumed mass of progenitors (see Stanghellini
\& Haywood 2010 for a detailed discussion of the different PN types).
The horizontal error bar for PNe indicates the  age range of the population, while all vertical error bars indicate
the uncertainties in the slopes. It appears that 
the slope of the M81 metallicity gradient did not  
vary very much during the past 10 Gyr. A flattening with time, as predicted by many chemical 
evolution models, is not supported by our observations. On the contrary, a constant slope, or a mild steepening of gradient slopes with time
is inferred.

\begin{figure} 
\resizebox{\hsize}{!}{\includegraphics{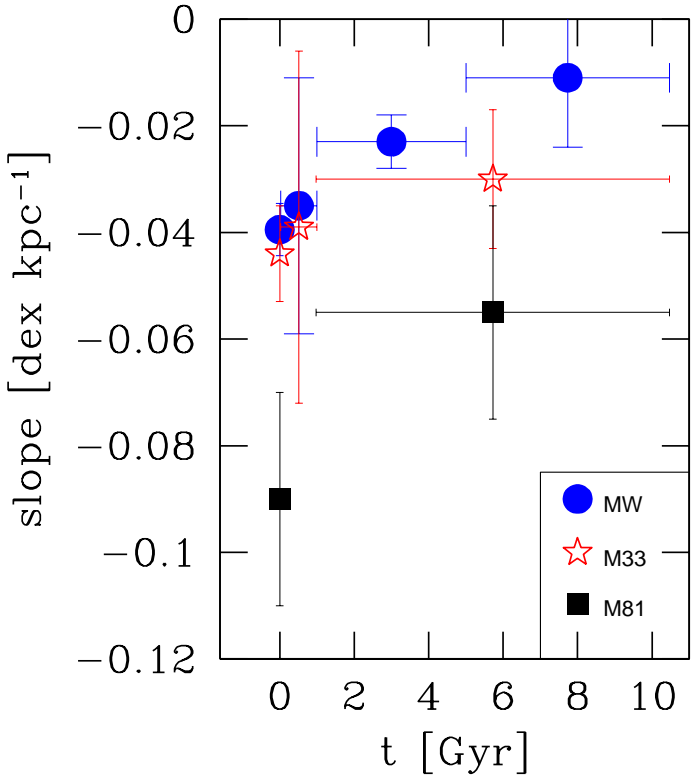}} 
\caption{The slope of the oxygen gradients, in dex kpc$^{-1}$, versus the age of the stellar 
population considered, for the Galaxy (filled circles, adapted from Stanghellini \&
Haywood 2010), M33 (starred symbols, adapted from M09 and M10) and M81 (squares, this paper). The zero-age populations represent the slopes of the gradients estimated 
though \hii~ region oxygen abundances, while all other points are slopes of
 PN gradients, see text.
Error-bars in the x direction represent the approximate age span of the populations, and in the
y direction are the estimated metallicity uncertainties. The zero-age data point for M81 is uncertain, as it is based on the combination of 
of the \hii\ regions observed with the MMT and the GS87 sample} 
\label{} 
\end{figure}

While uncertainties in the actual slopes are high, it is worth comparing the results of M81 with what found in M33 PNe and \hii~ regions 
(M09 and M10, starred symbols), and the Milky 
Way (PNe from Stanghellini \& Haywood 2010, \hii~ regions from Deharveng et al. 2000, filled circles). 
The metallicity gradients are generally steeper in M81
than in either the Milky Way and M33, but in all three galaxies we observe mild flattening with the age of the population, 
going from steeper for \hii~ regions to flatter for non-Type I PNe. It is worth noting that in the Galaxy the PN types I, 
II, and III were independently selected to calculate the oxygen radial gradients, while in
M33 we were able to distinguish between Type I and non-Type I. 

Since the gradient slopes seem to flatten with the age of the population considered, there is an indication that the gradients 
themselves are steepening with the galaxy age in all the three galaxies considered. It is a mild steepening, and with 
large uncertainties, but it is consistently found in 
all three galaxies, so it is worth exploring it. 

Let us recall that the slope of the \hii\ gradient in M33 excluded the central kpc, as explained by M10.
If we were including \hii~ regions in the central kpc we would obtain nearly no gradient evolution between \hii~ regions and Type I PNe in
M33. There is thus some evolution of the oxygen gradients with time, where the slopes vary by 1$\times10^{-2}$ to$\sim3\times10^{-3}$ dex per 
radial kpc in $\sim$10 Gyr. Stanghellini \& Haywood (2010) offer the hypothesis that portions of the 
Milky Way at different radial distances might have evolved at different rates. The model by Sch\"onrich \& Binney (2009) illustrates this type of evolution, 
which could hold as well for the other spiral galaxies and for M81. 

In Figure 8 we show the global oxygen evolution in different galaxies, where average oxygen abundances in \hii\ regions are plotted against
the same quantities in non-Type I PNe. In the case of the disk galaxies, both populations cover a reasonable 
radial range and are representative of the whole disks. The SMC and LMC oxygen abundance are from Leisy \& Dennefeld (2006), and the \hii~ regions 
in the Magellanic Clouds are from Dennefeld (1989). The samples of PNe and \hii~ regions considered in the Galaxy and M33 are as described earlier in this 
section. M81 PN and \hii\ regions abundances are from this paper, both selected consistently within the galactocentric distance domain where both sets of
targets coexist. The error-bars in the lower right of the plot
represent conservative data ranges for the galaxies considered here. The solar point is from Asplund et al. (2005). In this plot M81 stands out as the galaxy
with most chemical enrichment in the group considered. M33, as found by M10, shows almost no enrichment, and the other galaxies show either
low enrichment (the Galaxy and the LMC) or slight depletion of oxygen in the SMC, 
where the cause could be extremely efficient ON cycle 
at  very low metallicity (Karakas \& Lattanzio 2007). The average enrichment of M81 is 0.25 dex.

\begin{figure} 
\resizebox{\hsize}{!}{\includegraphics{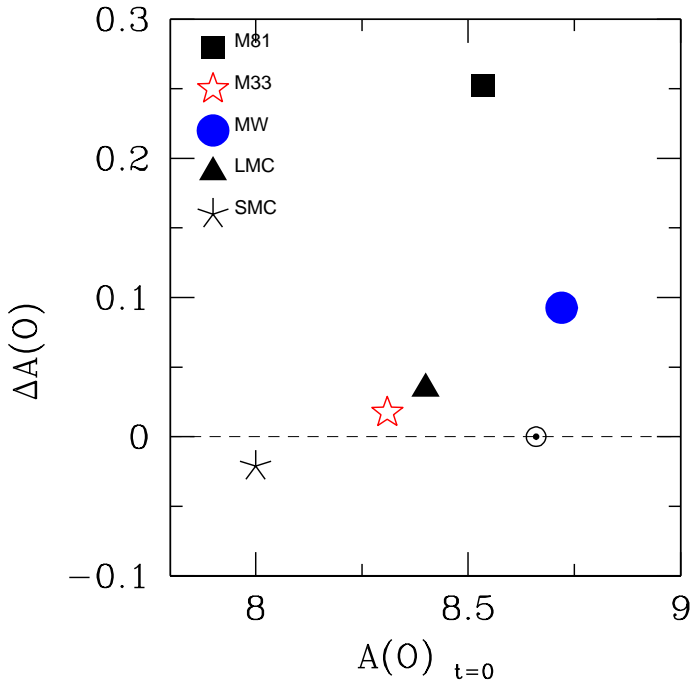}} 
\caption{Difference between average oxygen abundance of \hii~ regions and non-type I PNe vs. zero-age average oxygen abundance for M81 (square, this paper), compared to that of M33 (starred symbol, M09 and M10),
the Magellanic Clouds (LMC: triangle, SMC: diamond; Garnett 1999, M09), and the Milky Way (filled circle, Stanghellini \& Haywood 2010, Peimbert 1999). The solar value 
represented is from Asplund et al. (2005). The M81 datum is derived from PN and \hii\ regions in the 5.8-9.8 kpc range, and does not include the GS87 data.The M81 galaxy shows considerable metal enrichment.} 
\label{} 
\end{figure}

The M81 galaxy disk shows indication of chemical enrichment through oxygen abundances. It also shows generally steeper metallicity gradients than 
both M33 and the Galaxy. The non-type I PNe gradients in M81 is - 0.055 dex kpc$^{-1}$, while homologous slopes are - 0.03 and - 0.02 
dex kpc$^{-1}$ respectively in M33 and the Galaxy, indicating that whatever process produces the metallicity gradients in spiral galaxies, it is more
efficient in M81 and the overall galaxy metallicity is not a factor. Similarly for \hii~ regions, the M81 gradient slope is - 0.09 dex kpc$^{-1}$, to be compared with - 0.044 
and -0.04 dex kpc$^{-1}$ for M33 and the Milky Way respectively. Indication of steeper gradients, and sizable chemical evolution, are thus the 
defining characteristics of M81 compared with other spirals. The oxygen enrichment indicates that this galaxy has suffered outflow to a lesser extent than the
comparison galaxies, and steeper oxygen gradients are compatible with this explanation.

\setcounter{table}{6}
 \begin{table}
\caption{Properties of M81 and the Galaxy}
\begin{tabular}{lrr}
\hline\hline
								&M81						  & MW \\
\hline
Morphological type 		&Sa						  					&Sbc		\\
Total mass [M$_{\odot}$]				&2.3$\times$10$^{11}$(a) &1.8-3.7$\times$10$^{11}$ (b) \\
Gas mass  [M$_{\odot}$]					&2.9$\times$10$^9$(c)  	&4$\times$10$^9$ (d)\\
Luminosity  [L$_{\odot}$]		&1.9$\times$10$^{10}$ (a)& $\dots$\\
Total SFR	[M$_{\odot}$yr$^{-1}$]				&0.4-0.8 (e)				&0.35 (f)\\
Max V$_{rot}$	[km s$^{-1}$]			&260 (g)				& 200 (h)	 \\
\hline
\end{tabular}\\
(a) Appleton et al. (\cite{appleton81});
(b) Hodge \& Miller (\cite{hodge92});
(c) Karachentsev \& Kashibadze (\cite{kara06});
(d) van der Kruit (\cite{vdk90});
(e) Gordon et al. (\cite{gordon04});
(f) Walterbos \& Braun (1994);
(g) Rohlfs  \& Krertschmann (\cite{rohlfs80});
(h) MD05. 
\label{tab_m81_mw}
\end{table}

\subsection{The chemical evolution of M81: comparison with the Milky Way}

 It is evident from the above analysis that the metallicity gradient
slopes in the galaxies examined do not depend on the average galactic
metallicity:  M33 is metal poor (its metallicity is similar to that of
the LMC, Leisy \& Dennefeld 2006) and M81 is closer in metal contents to
the Galaxy. Since M81 is also closer to the Galaxy in stellar mass
content (see Table 7 for a direct comparison of the main properties of
these two galaxies), it makes sense to compare these two galaxies
directly, to explore the relations of the metallicity gradients to the
characteristics of  the galaxy disks and to their evolution.  The
observations of  PNe tell us that the gradient of
M81 is steeper than that of our Galaxy in old stellar populations; the trend 
seems to persist in the young stellar population, but this needs further confirmation.

From a theoretical point of view, there are two main reasons that could
influence the gradient slopes, (1) the rotational velocity of the
galaxy and (2) the different situation of these two galaxies within
their environment.

Moll\'a \& Diaz (2005) noted that metallicity gradients depend on the
rotational velocity of the galaxy (thus presumably on its total mass),
and on its morphological type: more massive galaxies tend to evolve
faster and have flatter gradients than lower mass galaxies; for a given
rotation velocity, gradients are steeper for late type than for early
type galaxies. Both effects would imply a steeper gradient for the
Galaxy than for M81, contrary to the observations. Some other factor
must be at play.

It is likely that the situation of M81 within its group of
galaxies influences its metallicity gradient.  The tidal \hi\ features
near M81 are consistent with a large-scale redistribution of gas in
this galaxy.  M81 is the only massive galaxy of its group, surrounded
by several small galaxies.  The interaction of M81 with the dwarf
galaxies of its group probably steepens the gradient of the major
member due to the effect of stripping gas from the external regions
of the major companion, while interaction of galaxies with more or less
the same mass would redistribute the gas in both galaxies, and thus
flatten the gradient.

The chemical evolution models of Valle et al. (2005) consider the
effects of gas stripping due to galaxy encounters on the star formation
rate and the evolution of the metallicity.  They found that, for a
stripping occurring 1-3 Gyr after the formation of the galaxy and
removing 97$\%$ of the gas, the region affected by the gas removal has
a SFR almost a factor of 10 lower than in the model without stripping
and the relative metallicity is then reduced by about 40$\%$.  The
metallicity reduction is not strongly dependent on the time and
duration of the stripping episode, but is quite sensitive to the
relative amount of gas removed from the region.  This effect should be
more pronounced in the outer than the inner galactic regions, due to
the proximity of the interacting galaxies, and it might explain why M81
has steeper metallicity gradients than the Galaxy. Unfortunately, the
models by Valle et al. (2005) are limited to a single radial region,
thus this conclusion is of qualitative nature.  A radially-resolved
chemical evolution modeling taking into account the effects of
stripping would be necessary to determine the effect of tidal
interaction on the gradient, and also to compare with other results.

\section{Conclusions}

Hectospec/MMT spectroscopy of a sizable sample of PN and \hii\ regions in the nearby M81 galaxy has proven very efficient to find chemistry of the 
young and old stellar populations, and to pin point the radial metallicity gradients. We were able to detect the diagnostic lines for plasma and 
abundance analysis in 19 PNe and 14 \hii\ regions. Their analysis indicates that the galaxy is clearly chemically enriched, with 
$\rm{<O/H>_{\hii\ reg.}/<O/H>_{PN}}$=1.8, from MMT spectra where PNe and \hii\ regions were simultaneously acquired. 
We also found that there is a noticeable PN metallicity gradient in oxygen, with
$\Delta{\rm log(O/H)}/\Delta{\rm R_G}$=-0.055 dex kpc$^{-1}$, and that
neon and sulfur gradient slopes are within 15$\%$ of the oxygen one. The MMT sample of \hii\ regions have limited galactocentric distribution, thus they are insufficient probes of the metallicity gradient. The gradient slope from the combined MMT and GS87 \hii\ region samples
is steeper (-0.093 dex kpc$^{-1}$) than that of the PNe, possibly indicating an evolution of the radial metallicity gradients with time: older stellar population
show shallower gradients, thus gradients are steepening with the time since galaxy formation. These results have been compared to their homologous
for the Milky Way and M33, where similar (yet less marked) gradient steepening is inferred.
We plan to increase the size of the \hii\ region sample, and to extend their gradient study to other atoms such as neon and sulfur in the near future, for
a more complete comparison with the PNe, with the goal of obtaining firmer observational constraints  for the chemical evolutionary models.


\begin{acknowledgements}
We warmly thank  D. Fabricant for making Hectospec available to the 
community, and the Hectospec instrument team and MMT staff for their expert 
help in preparing and carrying out the Hectospec observing runs. We thank N. 
Caldwell, D. Ming, and their team for the help during the data reduction. L.S. 
acknowledges the hospitality of the {\it Observatoire de Paris} and the  {\it
Osservatorio Astrofisico di Arcetri} for their hospitality during different stages of this project.

\end{acknowledgements}

\clearpage
\onecolumn
\setcounter{table}{3}

\begin{table}
\caption{Temperatures and Abundances Uncertainties}
\begin{tabular}{lrrrrrr}
\hline\hline
PN&$\Delta$F$_{\lambda}$&
$\Delta$\tenii& 
$\Delta$\teoiii& 
$\Delta$A(O)& 
$\Delta$A(N)& 
$\Delta$A(S)\\
&
$\%$&
$\%$&
$\%$&
dex&dex&dex\\
(1)& (2)& (3)& (4)& (5)& (6)& (7)\\

\hline

 PN~5 & 16&  		$\dots$& 	6&  0.09&  0.06&  0.06\\
 PN3~m  & 23& 	8& 	$\dots$& 0.17&  0.11&  0.10\\
 PN9~m & 27,5&  9& 2&  0.21&  0.13&  0.12\\
 PN158~m& 48&$\dots$& 23& 0.29& 0.20& 0.19\\

\hline
\end{tabular}

\end{table}

\begin{table}
\caption{Galactocentric distances and elemental abundances}
\begin{tabular}{lrrrrrrr}
\hline\hline
Name& R$_{\rm G}$ [kpc]& He/H& A(N)& A(O)& A(Ne)& A(S)& A(Ar)\\
(1)&(2)&(3)&(4)&(5)&(6)&(7)\\
\hline
    PN~3m &    10.49$\pm$0.177 & 0.051$\pm$0.002 	& 7.53$\pm$0.11 	& 8.59$\pm$0.17 &    $\dots$ & 6.85$\pm$ 0.10 & $\dots$  \\
     PN~5 &    6.276$\pm$0.094 &  $\dots$			& 7.41$\pm$0.06 	& 8.17$\pm$0.09 &   $\dots$ & 6.85$\pm$0.06 & 6.37$\pm$0.06 \\
    PN~9m &    7.344$\pm$0.156 & 0.101$\pm$0.002 	& 7.49$\pm$0.13 	& 8.54$\pm$0.21 & 7.67$\pm$0.12 & 6.51$\pm$0.12 & 5.76$\pm$0.12 \\
   PN~15m &    9.006$\pm$0.007 &    $\dots$ 			&     $\dots$		& 8.18$\pm$0.16 &    $\dots$&    $\dots$ &    $\dots$ \\
   PN~17m &    4.858$\pm$0.261 & 0.138$\pm$0.015 	& 7.59$\pm$0.13 	& 8.59$\pm$0.21 &    $\dots$& 6.87$\pm$0.12 &     $\dots$ \\
   PN~29m &    12.14$\pm$0.415 & 0.109$\pm$0.017 	& 7.13$\pm$0.05 	& 8.24 $\pm$0.06 & 7.71$\pm$0.05 & 6.41$\pm$0.05 &    $\dots$\\
   PN~33m &    10.86$\pm$0.110 &  $\dots$			 & 7.07$\pm$0.18 	& 7.94$\pm$0.28 &    $\dots$ & 6.43$\pm$0.18 &    $\dots$\\
   PN~38m &    12.54$\pm$0.385 & 0.122$\pm$0.061 	&     $\dots$		& 8.11$\pm$0.22 &     $\dots$ &    $\dots$ & 6.08$\pm$0.13 \\
   PN~41m &    4.677$\pm$0.110 & 0.179$\pm$0.063 	& 7.20$\pm$0.18 	& 8.11$\pm$0.28 &    $\dots$ & 6.78$\pm$0.18 &     $\dots$\\
   PN~45m &    4.326$\pm$0.018 & 0.078$\pm$0.006 	& 7.84$\pm$0.06 	& 8.89$\pm$0.09 &    $\dots$& 6.95$\pm$0.06 &     $\dots$ \\
   PN~63m &    3.516$\pm$0.117 & 0.089$\pm$0.002 	& 7.74$\pm$0.14 	& 8.62$\pm$0.22 &     $\dots$& 6.86$\pm$0.13 &    $\dots$\\
   PN~70m &    4.785$\pm$0.084 & 0.127$\pm$0.005 	& 7.67$\pm$0.18 	& 8.73$\pm$0.28 &    $\dots$ & 6.97$\pm$0.18 &     $\dots$ \\
   PN~93m &    2.805$\pm$0.109 & 0.068$\pm$0.003 	& 7.79$\pm$0.05 	& 8.51$\pm$0.05 & 8.06$\pm$0.05 & 7.10$\pm$0.05 &  $\dots$\\
  PN~121m &    4.119$\pm$0.218 & 0.184$\pm$0.014 	& 7.66$\pm$ 0.3 	& 8.81$\pm$ 0.50 &   $\dots$& 6.98$\pm$ 0.30 &$\dots$ \\
  PN~128m &    5.339$\pm$0.265 &   $\dots$ 			& 7.83$\pm$0.18 	& 8.27$\pm$0.28 &    $\dots$& 7.16$\pm$0.18 &    $\dots$ \\
  PN~149m &    10.14$\pm$0.547 & 0.145$\pm$0.018 & 7.60$\pm$0.04 	& 8.13$\pm$0.04 & 7.62$\pm$0.04 & 6.48$\pm$0.04 & 5.84$\pm$0.04 \\
  PN~153m &    8.381$\pm$0.277 & 0.147$\pm$0.007	&7.40$\pm$0.13 	& 8.31$\pm$0.21 & 8.04$\pm$0.12 & 6.78$\pm$0.12 &    $\dots$\\
  PN~157m &    9.759$\pm$0.493 & 0.201$\pm$0.023 	&   $\dots$ 		& 8.07$\pm$0.04 &    $\dots$ &   $\dots$&   $\dots$\\
  PN~158m &    10.15$\pm$0.525 & 0.166$\pm$0.078 	&  7.37$\pm$ 0.20	& 7.94$\pm$0.29 & 7.40$\pm$0.19 & 6.91$\pm$0.19 & $\dots$\\
&&&&&&&\\
Average& &0.127$\pm$0.045&  7.58$^{+0.18}_{-0.30}$&   8.45$^{+0.23}_{-0.54}$& 7.81$^{+0.20}_{-0.38}$& 6.86$^{+0.17}_{-0.28}$& 6.08$^{+0.22}_{-0.48}$\\
\hline
       HII~4 &    9.076$\pm$0.396 & 0.079$\pm$0.034	& 7.15$\pm$0.33 	& 8.58$\pm$ 0.50 & 7.92$\pm$0.34 & 6.45$\pm$0.33 & 6.11$\pm$0.33 \\
       HII~5 &    8.696$\pm$0.342 & 0.072$\pm$0.021	& 8.06$\pm$0.07 	& 8.87$\pm$0.04 &   $\dots$ & 6.63$\pm$0.03 & 5.87$\pm$0.03 \\
      HII~31 &    8.648$\pm$0.455 & 0.079$\pm$0.026	& 7.60$\pm$0.32 	& 8.60$\pm$ 0.50 &    $\dots$& 6.87$\pm$0.32 & 6.01$\pm$0.32 \\
        HII~72 &    6.773$\pm$0.070 & 0.086$\pm$0.011	& 7.87$\pm$0.13 	& 8.82$\pm$0.22 & 8.28$\pm$0.13 & 6.79$\pm$0.13 & 6.05$\pm$0.13 \\
      HII~79 &     8.16 0$\pm$0.006 & 0.077$\pm$0.020	& 7.30$\pm$0.29 	& 8.26$\pm$0.45 & 7.62$\pm$0.29 & 6.63$\pm$0.29 & 5.88$\pm$0.29 \\
      HII~81 &    7.027$\pm$0.031 & 0.104$\pm$0.037	& 7.41$\pm$0.32 	& 8.40$\pm$ 0.50 & 7.73$\pm$0.32 & 6.59$\pm$0.32 & 5.89$\pm$0.32\\
       HII~123 &    7.762$\pm$0.002 & 0.084$\pm$0.007 	& 7.40$\pm$0.10 	& 8.43$\pm$0.18 & 7.68$\pm$0.10 & 6.76$\pm$0.10 & 6.30$\pm$0.10 \\
     HII~133 &    6.721$\pm$0.001 & 0.091$\pm$0.159 	& 7.90$\pm$0.05 	& 8.67$\pm$0.04 & 8.19$\pm$0.05 & 6.70$\pm$0.05 & 6.04$\pm$0.05 \\
     HII~201 &     6.750$\pm$0.070 & 0.107$\pm$0.032 	& 7.63$\pm$0.13 	& 8.26$\pm$0.22 & 7.73$\pm$0.13 & 6.55$\pm$0.13 & 5.97$\pm$0.13 \\
     HII~228 &    9.849$\pm$0.211 & 0.052$\pm$0.021 	& 7.41$\pm$0.04 	& 8.35$\pm$0.04 & $\dots$& 6.79$\pm$0.04 & 5.92$\pm$0.04 \\
     HII~233 &    5.797$\pm$0.138 & 0.086$\pm$0.043 	& 7.86$\pm$0.25 	& 8.50$\pm$ 0.38 & $\dots$&  6.54$\pm$0.25 & $\dots$\\
     HII~262 &    9.716$\pm$0.280 & 0.073$\pm$0.017 	& 7.33$\pm$0.07 	& 8.32$\pm$0.09 &$\dots$ & 6.73$\pm$0.07 & 6.06$\pm$0.07 \\
     HII~325 &    9.261$\pm$0.403 & 0.081$\pm$0.042	& 7.23$\pm$0.37 	& 8.18$\pm$ 0.60 &  $\dots$& 6.42$\pm$0.37 & 5.96$\pm$0.37 \\ 
     HII~403 &    9.653$\pm$0.493 & 0.069$\pm$0.016	& 7.46$\pm$0.13 	& 8.57$\pm$0.22 &  $\dots$ & 6.87$\pm$0.13& 6.21$\pm$0.13\\
 &&&&&&&\\ 
Average&   & 0.082$\pm$0.013& 7.63$^{+0.23}_{-0.54}$& 8.54$^{+0.18}_{-0.32}$& 7.95$^{+0.22}_{-0.47}$& 6.69$^{+0.12}_{-0.17}$& 6.04$^{+0.13}_{-0.18}$\\
\hline
\end{tabular}
\end{table}


\end{document}